\documentclass[journal,twoside,web]{ieeecolor}
\usepackage{generic}
\usepackage{cite}
\usepackage{amsmath,amssymb,amsfonts}
\usepackage{algorithmic}
\usepackage{graphicx}
\usepackage{textcomp}
\def\BibTeX{{\rm B\kern-.05em{\sc i\kern-.025em b}\kern-.08em
    T\kern-.1667em\lower.7ex\hbox{E}\kern-.125emX}}
\begin{document}
\title{SHE-MTJ Circuits for Convolutional Neural Networks}
\author{Andrew W. Stephan and Steven J. Koester, \IEEEmembership{Fellow, IEEE}
\thanks{Manuscript submitted \today. This work was supported by Seagate Technology PLC.}
\thanks{The authors acknowledge the Minnesota Supercomputing Institute (MSI) at the University of Minnesota for providing resources that contributed to the research results reported within this paper. URL: http://www.msi.umn.edu}
\thanks{The authors wish to thank Chris Kim and Minsu Kim with the University of Minnesota and Qiuwen Lou with the University of Notre Dame for helpful discussions.}
\thanks{A. W. Stephan is with the College of Science and Engineering, University of Minnesota, Minneapolis, MN 55455 USA (e-mail:steph506@umn.edu).}
\thanks{S. J. Koester is with the College of Science and Engineering, University of Minnesota, Minneapolis, MN 55455 USA (e-mail:skoester@umn.edu).}}

\maketitle

\begin{abstract} We report the performance characteristics of a notional Convolutional Neural Network based on the previously-proposed Multiply-Accumulate-Activate-Pool set, an MTJ-based spintronic circuit made to compute multiple neural functionalities in parallel.  A study of image classification with the MNIST handwritten digits dataset using this network is provided via simulation. The effect of changing the weight representation precision, the severity of device process variation within the MAAP sets and the computational redundancy are provided. The emulated network achieves between 90 and 95\% image classification accuracy at a cost of ~100 nJ per image.
\end{abstract}


\section{Introduction}
\label{sec:introduction}

Convolutional neural networks (CNNs) are a powerful tool for beyond-Boolean computing such as data classification, whether it be text, audio or visual.\cite{CoNNs, CeNNs with CoNNs} Their complexity makes in-hardware implementations difficult and costly beyond that of the simpler fully-connected network. We propose to use the Multiply-Accumulate-Activate-Pool (MAAP) sets described in \cite{MAAP} to reduce the complexity of convolutional neural network implementation. The MAAP sets will limit the number of unique operations required by the CNN by condensing the convolution, activation and pooling operations into one circuit. In so doing, the number of required peripheral operations such as memory is also reduced compared to other hardware-based implementations that incorporate all CNN functions individually\cite{CeNNs with CoNNs}.

\section{Background}
\label{sec:Background}

\begin{figure*}
\centering
\includegraphics[scale=0.53]{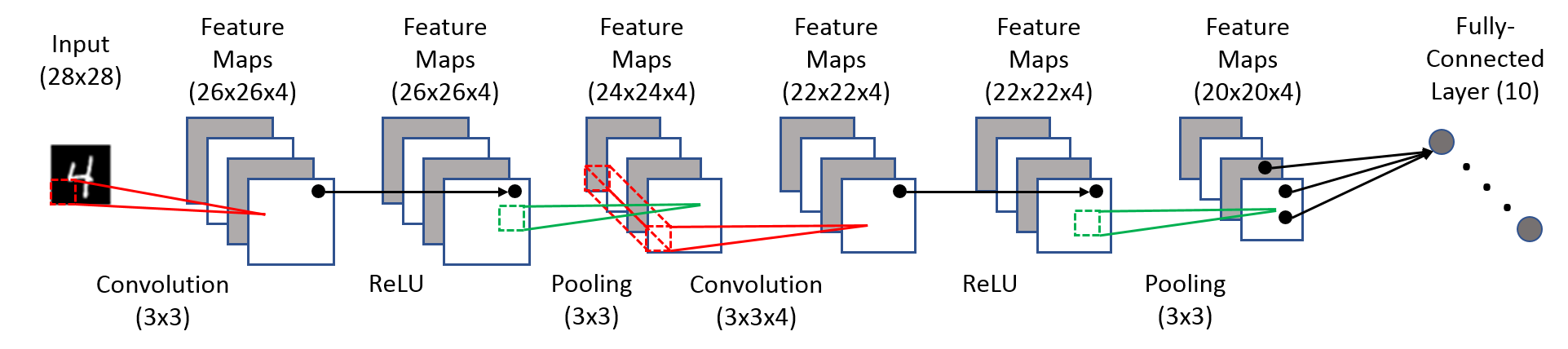}
\caption{The CNN structure used for training in TensorFlow. The network consists of two convolutional layers with their subsequent activation and pooling layers, followed by one final fully-connected layer. Each convolution contains four kernels.}
\label{fig:CNN}
\end{figure*}

\subsection{Convolutional Neural Networks}
Typical CNNs consist of one or more sequences of convolution, activation and pooling layers followed by one or more fully-connected (FC) layers as shown in Fig. \ref{fig:CNN}. Each neuron in the convolutional layers applies a certain weight template to a subset of values in the input space, with neighboring neurons applying the same template to neighboring, possibly overlapping, subsets. Since multiple values in the input contribute to a single value in the convolution layer, this results in down-sampling of the image size. Each template corresponds to one full  set of neurons, or one convolutional image map. Activation layers pass each value contained in the convolutional layers through some nonlinear activation function in a 1-to-1 fashion. The rectified linear unit (ReLU) 
\begin{gather}
R(x) = 
\begin{cases}
0 & \text{for } x < 0\\
x & \text{for } x \ge 0
\end{cases}
\label{eq:ReLU}
\end{gather}
is commonly used for this purpose. In max-pooling, each neuron chooses the maximum value from its unique subset of the input space. This further down-samples the data and also introduces some translation-invariance. The fully-connected layer consists of a one-dimensional vector of neurons, each of which takes a weighted sum of all values in the previous layer. The convolution, activation and pooling layers prior to the final fully-connected layer comprise a significant portion of the computational cost of a CNN. 

\subsection{MAAP Sets}
The spin-torque-controlled magnetic tunnel junction (MTJ) is a well-known basic element of spintronic computing.\cite{SHE MTJs,KRoy,mLogic,Naeemi1,Naeemi2} Depending on the circuit layout, geometry and specific application of spin torque, these versatile spin-MTJs can be used as analog or digital programmable synapse memristors, spiking neurons or artificial neurons. In \cite{MAAP} a useful application of MTJ cells manipulated via the spin-Hall effect (SHE) is proposed. Utilizing in-plane fieldlike spin-torque along the hard axis of the free layer (FL), a linear hysteresis loop is produced\cite{Fieldlike}. With the appropriate choice of circuit parameters, a voltage divider composed of one SHE-MTJ cell and a reference resistor with an inverter to read the output can produce a linear output--with saturation--as a function of the charge current passing through the SHE layer. This structure is effectively a three-terminal device with input, output and constant terminals in which the potential across the output and constant terminals depends on the charge current injected from the input terminal to the constant terminal. Crossbar arrays are commonly used in neuromorphics to perform the multiply-and-accumulate operation by summing up parallel currents, each of which represents a single product of a voltage and a conductance value. In order for the sum to be correct, leakage between the parallel lines must be minimized by holding the bottom potential constant. In order to inject this current sum as an input to a device with low error, the device must have very low input impedance so that the floating potential on the input terminal is very close to the value on the constant terminal regardless of the actual input. The SHE-MTJ voltage divider stack uniquely accomplishes this by using a low-resistance SHE layer to read the charge current and transform it into a spin signal without significantly disturbing the input potential. An equivalent circuit based entirely on charge signals would require additional amplifiers at greater cost to maintain the input terminal at a constant potential. A ReLU activation pair consists of two concatenated SHE-MTJ cells and readout inverters with additional bias current sources. Several such activation pairs may be organized into a winner-take-all circuit that simultaneously selects the maximum of a set of input values and computes the ReLU activation on the input. This circuit was shown to compute efficiently while also being robust to both thermal and process variation, including MTJ resistance state variation, critical-current variation and transistor threshold variation\cite{MAAP}. This makes it a good candidate for a spintronic CNN accelerator.

\section{MAAP-CNNs}
\label{sec:MAAP-CNNs}

\begin{figure}
\centering
\includegraphics[scale=0.45]{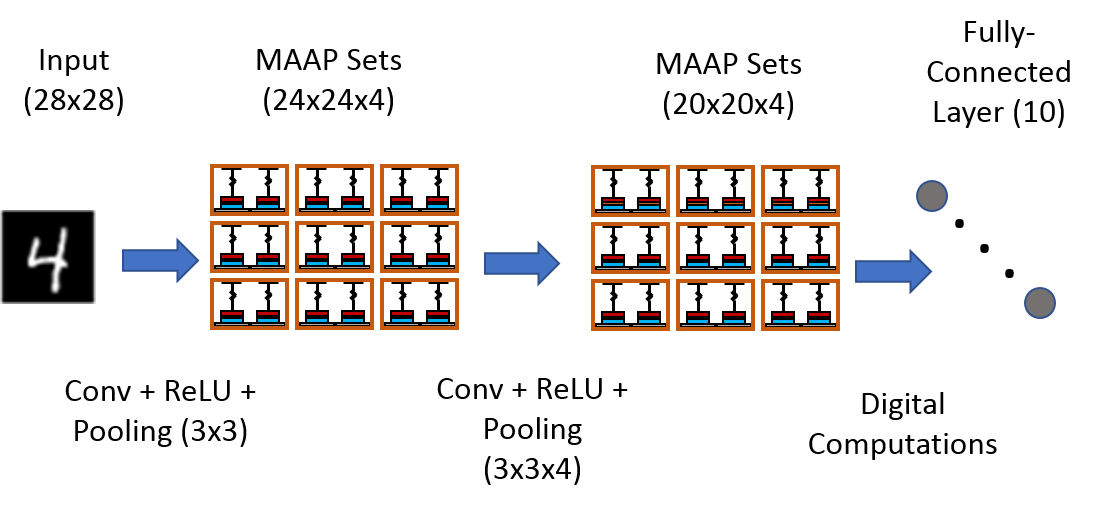}
\caption{The decomposition of the CNN in Fig. \ref{fig:CNN} to a MAAP-CNN. Each convolution-activation-pooling layer sequence has been replaced by a matching group of MAAP sets which perform all three functions.}
\label{fig:MAAPCNN}
\end{figure}

\begin{figure}
\centering
\includegraphics[scale=0.45]{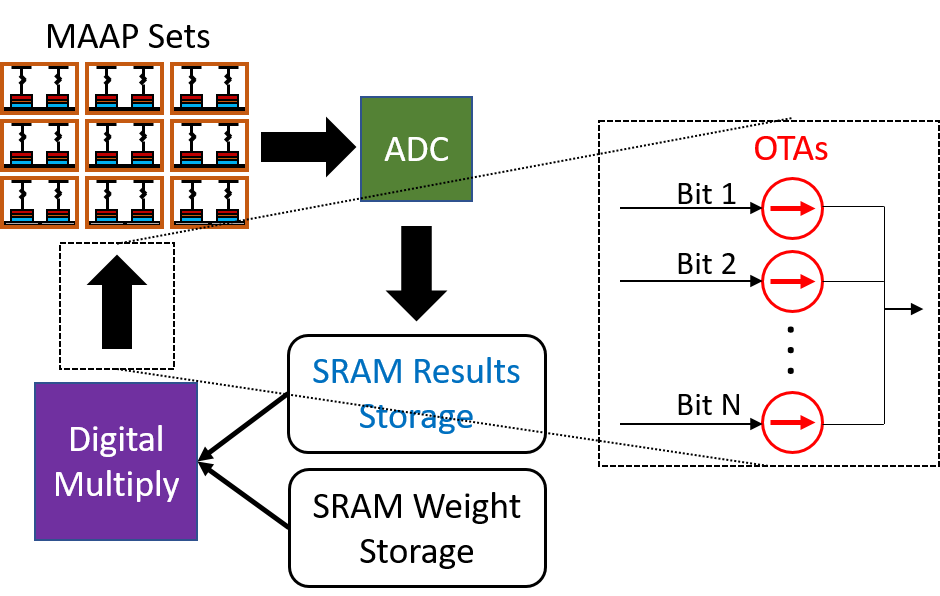}
\caption{Block diagram of MAAP set and memory. The weights and inputs are digitally stored in SRAM. Each MAAP output is converted via ADC and similarly stored. Weight multiplication is accomplished digitally before converting back to analog current via a crossbar-like array of OTAs.}
\label{fig:Layout}
\end{figure}

We applied the MAAP set concept to the problem of classifying the MNIST handwritten digits dataset. The HSPICE/Matlab simulator used in \cite{MAAP,PTM} was used to generate a simplified input-to-output simulation module, vastly decreasing the required compute time. This module is used in Matlab to emulate the CNN in Fig. \ref{fig:CNN} as shown in Fig.\ref{fig:MAAPCNN} using weight templates trained in Tensorflow. To store intermediate results in the course of CNN processing the MAAP data is sent to analog-to-digital converters (ADCs) and stored in static-RAM (SRAM). The convolutional templates are also digitally stored in SRAM. These quantities are digitally multiplied, requiring approximately $B^2$ gates for $B$ bits, and supplied to $B$ operational transconductance amplifier (OTA) based current sources that provide the weighted input to the MAAP sets in a fashion similar to the OTA usage in \cite{MAAP,Qiuwen,OTA}. The low input impedance of the MAAP set circuit makes summing up parallel inputs in a crossbar-like manner quite accurate\cite{MAAP}. A block diagram of this layout is shown in Fig. \ref{fig:Layout}. We note that with sufficiently large memory, only one MAAP set is needed to emulate an entire CNN. However, since the processing time for a single MAAP operation is on the order of ns\cite{MAAP} it is much more time-efficient to complete the $\sim$4000 operations necessary with many sets(see Fig. \ref{fig:MAAPCNN}). 

\vspace{.5in}
\section{Results and Discussion}
\label{sec:Results}

\begin{figure}
\centering
\includegraphics[scale=0.47]{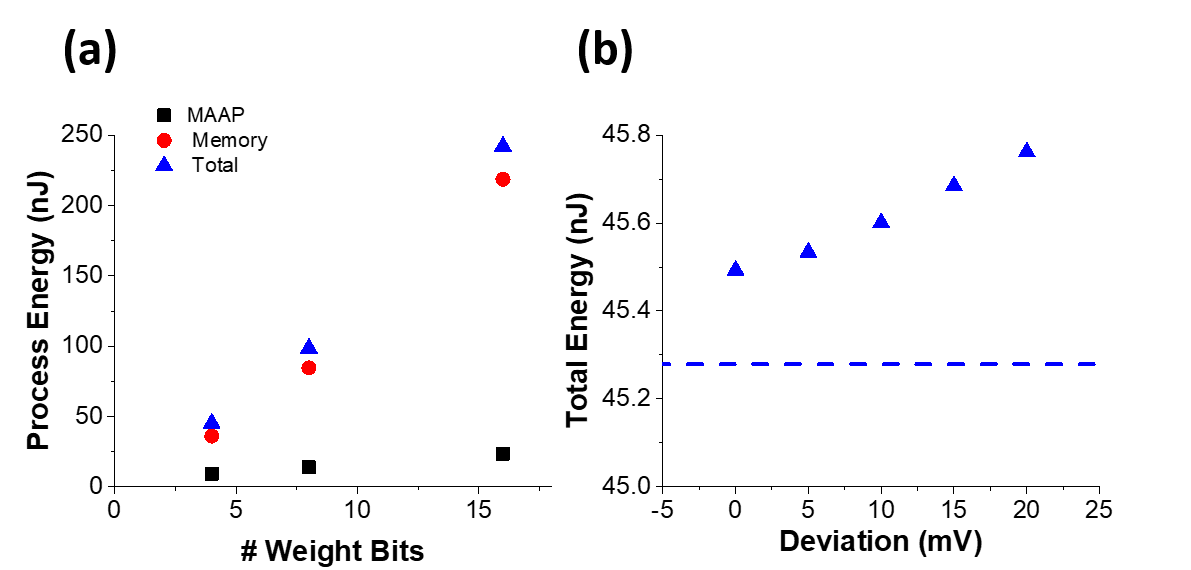}
\caption{(a) Energy consumed in the MAAP processing and memory operations vs. the number of bits used for each value. This data assumes ideal devices. (b) Total energy for non-ideal devices with different levels of transistor threshold deviation and constant FM parameter deviation. The dashed line indicates ideal device energy. The change is very small.}
\label{fig:Energy1}
\end{figure}

\begin{figure}
\centering
\includegraphics[scale=0.47]{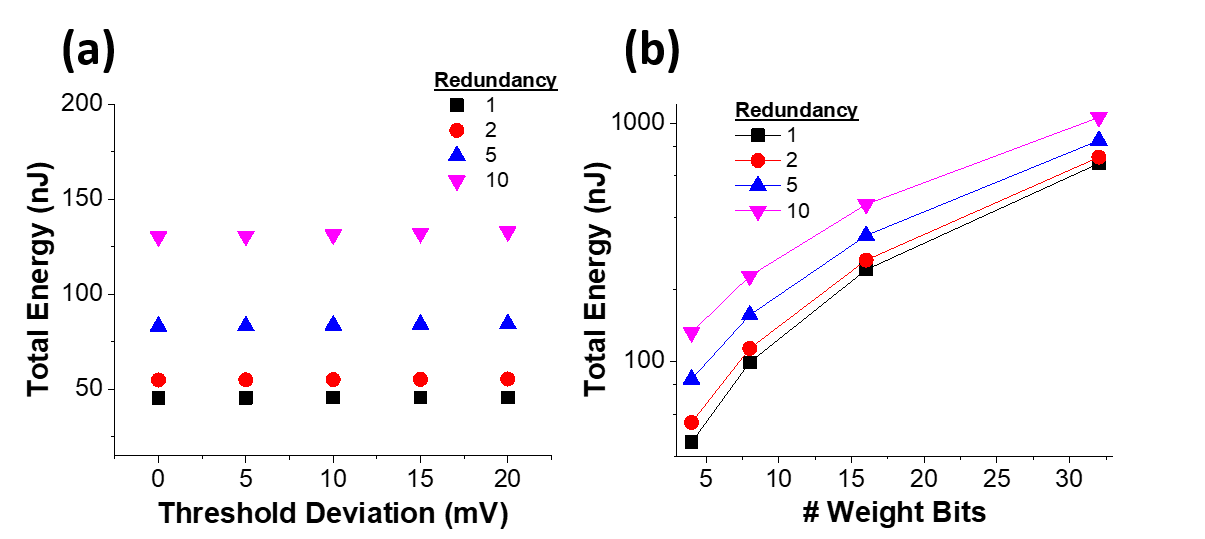}
\caption{(a) Total energy vs. the weight precision for different levels of threshold deviation. (c) Total energy vs. MAAP set redundancy for different levels of threshold deviation.}
\label{fig:Energy2}
\end{figure}

\begin{figure}
\centering
\includegraphics[scale=0.45]{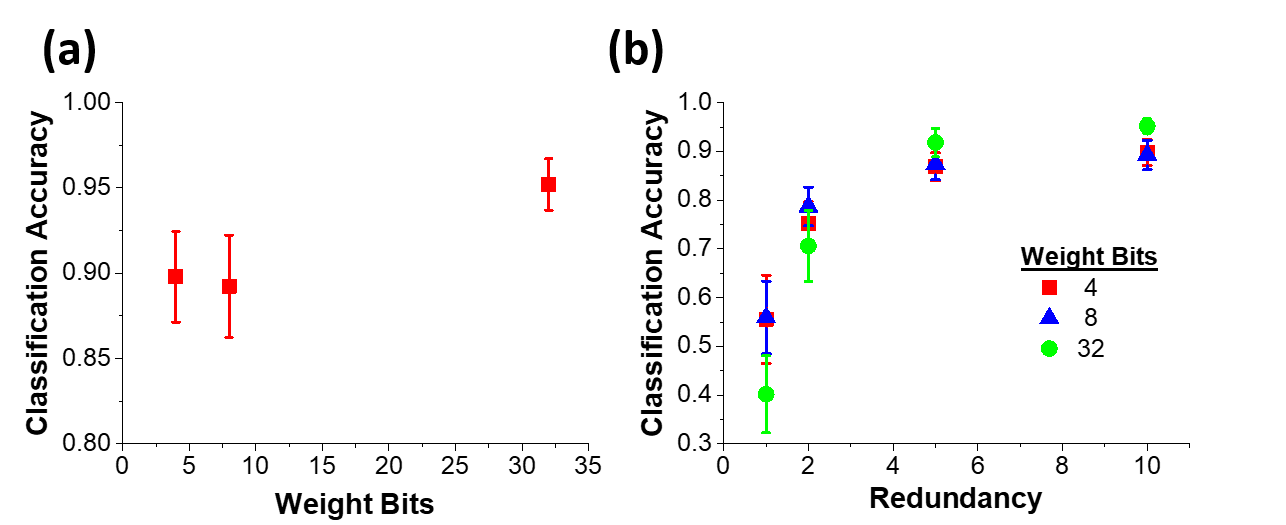}
\caption{(a) Classification accuracy on MNIST handwritten digit images vs. the weight representation precision. This plot assumes non-ideal devices with 15 mV of transistor threshold deviation and $R = 10$. (b) Classification accuracy vs. MAAP set redundancy using non-ideal devices with weight representation precision as a parameter. We note that at maximum redundancy the deleterious effect of process variation is negligible.}
\label{fig:Accuracy1}
\end{figure}

\begin{figure}
\centering
\includegraphics[scale=0.5]{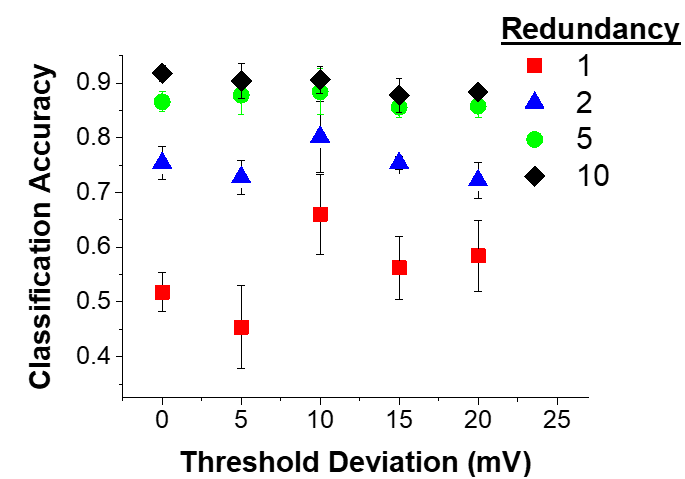}
\caption{Classification accuracy vs. transistor threshold deviation with redundancy $R$ as a parameter. All devices incorporate MTJ variation at a constant level consistent with the deviations in \cite{MAAP}. Only transistor threshold deviation severity is varied. We note that the accuracy appears not to vary with the particular amount of threshold deviation once any notable level of device variation is introduced.}
\label{fig:Accuracy2}
\end{figure}

\begin{figure}
\centering
\includegraphics[scale=0.6]{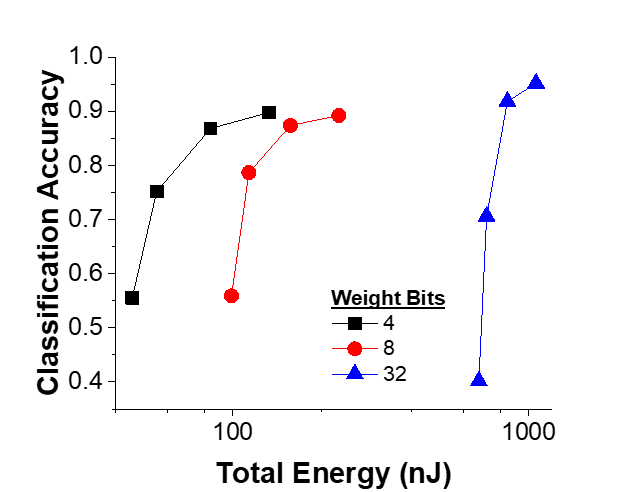}
\caption{Accuracy vs. energy for 4 bit, 8 bit and 32 bit networks. Negligible incremental benefit is granted for moving from 4 to 8 bit, but at 32-bit the final accuracy is increased by about 5 \%.}
\label{fig:AccvsEnergy}
\end{figure}

In this section we report the relationship between terms such as the image classification accuracy, redundancy factor $R$, bit representation precision $B$, energy and process variation. In the figures referenced herein, the voltage deviation term refers to threshold deviation. The other process variation terms related to the MTJs are assumed to follow constant Gaussian distributions with values drawn from those distributions for each device in each iteration. The width of the Gaussian threshold distribution is a variable in some figures.

\subsection{Energy}

\begin{table}
\centering
\caption{Operation Count}
\setlength{\tabcolsep}{3pt}
\begin{tabular}{|p{50pt}|p{50pt}|p{50pt}|}
\hline
Operation&
Energy/Op&
Iterations\\
\hline
\vspace{0.005in}
MAAP& 
\vspace{0.005in}
0.9 - 7.2 pJ&
\vspace{0.005in}
3904$\cdot R$\\
Memory&
41.8 fJ&
85328$\cdot B$\\
ADC&
1 pJ&
3904$\cdot B$\\
Multiply&
125.4 fJ&
3088$\cdot B^2$\\
\hline
\multicolumn{3}{p{150pt}}{Quantity of each operation needed to classify an image. $B$ and $R$ are the number of bits used for representation and the MAAP operation redundancy, respectively.}\\
\end{tabular}
\label{tab:opcount}
\end{table}

Apart from peripheral circuitry, the energy dissipation is largely independent of the number of physical MAAP circuits, being dependent on the number of MAAP and memory operations to be computed instead. The total energy dissipated during the processing of one image is:
\begin{gather}
E = N_{M}E_{M} + N_{MEM}E_{MEM} + N_{A}E_{A} + N_{MUL}E_{MUL},
\end{gather}
where $N_{M}$, $N_{MEM}$, $N_{A}$ and $N_{MUL}$ are the number of MAAP, digital memory, ADC and digital multiply operations, respectively. These quantities are given in Table \ref{tab:opcount}, with the values ultimately deriving from the dimensions of the MAAP set arrays shown in Fig. \ref{fig:MAAPCNN}. The energy terms $E_{M}$, $E_{MEM}$, $E_{A}$ and $E_{MUL}$ are the costs of a single instance of each operation. The MAAP operation cost $E_{M}$ accounts for leakage across the MTJ stacks, as calculated in\cite{MAAP} assuming a TMR of 1.5 and $40x40$ nm$^2$ MTJs, as well as the static and dynamic dissipation of the OTAs\cite{OTA}. $E_{MEM}$ is based on 16nm node transistor data for 6T SRAM.\cite{SRAM} $E_{MUL}$ is estimated with the same transistor data, assuming $B^2$ gates per $B$-bit multiplication. The ADC energy is taken from the Stanford ADC Survey, with a power to Nyquist sampling frequency ratio of 1 pJ.\cite{Stanford ADC Survey} We note that the majority of the total energy is taken up by the digital processing (see Fig. \ref{fig:Energy1}). The energy usage depends upon the level of precision used to encode the weights and store the MAAP outputs. The assumed level of process variation that needs to be corrected and the MAAP circuit runtime have some effect as well.\cite{MAAP} Fig. \ref{fig:Energy1} (a) compares the MAAP and memory operational costs for 4-bit, 8-bit and 32-bit networks, assuming 0.6 ns runtime per MAAP set. The memory and digital operations dominate the MAAP processing cost. Fig. \ref{fig:Energy1} (b) shows the rise in total energy required to correct the error in the mean MAAP output due to process variation, assuming a 4-bit network. The dashed line assumes no variation, while the data points all correspond to a set level of MTJ variation with standard deviations matching those reported in \cite{MAAP,Variation1,Variation2,Variation3}. The standard deviation for the transistor threshold potentials is varied with multiple values. Although the process variation-induced mean error is easily dealt with, the process variation introduces a significant deviation in the error as well. In order to reduce this effect, redundant sampling is used. Each MAAP set output is measured $R$ times with the same inputs and the final stored value is the average of the measurements. The variation in the expected mean of $R$ redundant measurements is less than the variation of a single measurement by a factor of $\sqrt{R}$. Multiplying the number of MAAP operations by the redundancy factor $R$ comes at the cost of additional energy for MAAP sampling, and the need for at least $R$ physically distinct MAAP sets as the different measurements must be performed on different devices to ensure a unique random sampling each time. Fig. \ref{fig:Energy2} shows the increase in total energy caused by operation redundancy.

\subsection{Accuracy}

In Fig. \ref{fig:Accuracy1} we display the results pertaining to image classification accuracy. The values shown are statistical estimates. Each simulation included 100 test images and the values shown in the figure are the mean result of five or more simulations with error bars included. Fig. \ref{fig:Accuracy1}(a) indicates the accuracy vs. $B$ using non-ideal simulated devices with redundancy $R = 10$ to indicate the upper bound on performance. As the representation precision grows the accuracy increases from an average of about 90\% to 95\%. Fig. \ref{fig:Accuracy1}(b) shows the accuracy vs. $R$. With $R = 1$, the high error deviation introduced by process variation significantly lowers accuracy; however, with a modest $R = 5$ the accuracy is almost entirely recovered. At $R = 10$ the results are indistinguishable from those of ideal devices. Surprisingly, the particular level of transistor threshold deviation appears insignificant to the accuracy as indicated in Fig. \ref{fig:Accuracy2}. The results do not significantly differ between devices with 0 mV or 20 mV of threshold deviation. The existence of any significant process variation--in the MTJs, if not the transistors--is sufficient to necessitate some redundancy. However, increasing the level of variation has little to no incremental effect, although we surmise that extreme amounts beyond what was tested would cause further noticeable deterioration of the accuracy. 

Finally, Fig. \ref{fig:AccvsEnergy} indicates the classification accuracy vs. the energy cost for three different bit precision levels. Using a 4-bit network with $R = 5$ is sufficient to reach nearly 90\% accuracy at a cost of about 80 nJ per image. The incremental cost of increasing accuracy to 95\% is quite large, requiring 32-bit precision and $R = 10$ with a cost of about 1000 nJ per image.

\subsection{Discussion}

This work demonstrates that spintronic circuits based on the increasingly well-understood MTJ neuron model can be used to effectively implement the complex functions involved in CNNs and accurately classify images. The MAAP architecture condenses the many different CNN layers into fewer, and more cohesive sets of calculations. The modularity of the MAAP set model also detaches the physical circuitry from the number of conceptual neurons involved in the CNN model. Depending on $R$ and $N$ the energy required to process an image varies between about $40 - 1000$ nJ, comparable to the networks reviewed in \cite{CeNNs with CoNNs}. At a maximum accuracy of between 90 - 95\% the performance is also comparable. We also note that, assuming sufficient devices to compute an entire sequence of layer operations in parallel, the fully charge-based network in \cite{CeNNs with CoNNs} requires between 101.5 and 139 ns to calculate the output of a single convolution-ReLU-pooling sequence. By reducing the number of times the network must pass its intermediary outputs through the ADC to memory and back, we achieve a significant speed-up. Each MAAP set operation takes on the order of 1 ns, saving a great deal of time if we assume identical delays in the ADC and digital processing peripherals between the two networks. Finally, comparing the number of MAAP operations $N_M = 3904$ to the number of CeNN operations in an equivalent network in \cite{CeNNs with CoNNs,IRMENCONN} $N_C \sim 56000$ shows a significant reduction in complexity, even with redundancy. We also note that condensing the neuromorphic layer operations reduces the number of ADC and memory operations that are required to store and access intermediate data compared to a system which explicitly computes each operation, especially those which compute the operations via multiple sub-steps which themselves may require digital processing. This should yield additional savings.  We hope this achievement will help spark continued interest and study of spintronic materials and circuits to unlock their great potential.

\end{document}